\documentclass{aa}
\usepackage{graphics}
\usepackage{txfonts}
\newcommand{\noi}{\noindent}
\def\bc{\begin{center}}
\def\ec{\end{center}}
\def\be{\begin{equation}}
\def\ee{\end{equation}}
\def\bea{\begin{eqnarray}}
\def\eea{\end{eqnarray}}
\def\ve{\varepsilon}
\def\mkm{\mu{\rm m}}
\def\ve{\varepsilon}
\def\bet{{\beta}~\rm Pictoris}
\begin{document}
\title{Dust extinction and absorption: the challenge of porous grains}
\titlerunning{Porous interstellar grains}

\author{ N.V.~Voshchinnikov\inst{1,2},
         V.B.~Il'in\inst{1,2},
         Th.~Henning\inst{3},
         and
         D.N.~Dubkova\inst{1}
         }
\authorrunning{Voshchinnikov et al.}

\institute{
Sobolev Astronomical Institute,
St.~Petersburg University, Universitetskii prosp. 28,
           St.~Petersburg, 198504 Russia,
e-mail: {\tt nvv@astro.spbu.ru}
\and
 Isaac Newton Institute of Chile, St.~Petersburg Branch
\and
Max-Planck-Institut f\"ur Astronomie, K\"onigstuhl 17, D-69117 Heidelberg, Germany
}
\date{Received 5 May 2005 / accepted 9 August 2005}

  \abstract{
In many models of dusty objects in space
the grains are assumed to be composite or fluffy.
However, the computation of the optical properties of such particles is still
a very difficult problem. We analyze
how the increase of grain porosity influences basic features
of cosmic dust ---
interstellar extinction, dust temperature, infrared bands and
millimeter opacity.

 It is found that an increase of porosity leads to an increase of
extinction cross sections at some wavelengths and a  decrease at others
depending on the grain model.
However, this behaviour is sufficient to reproduce the extinction curve in
the direction of the star $\sigma$ Sco using current solar abundances.
In the case of the star $\zeta$ Oph our model requires larger amounts
of carbon and iron in the dust-phase than is available.
Porous grains can reproduce
the flat extinction across the $3 - 8\,\mkm$ wavelength range measured
for several lines of sight by {\it ISO} and {\it Spitzer}.

Porous grains are generally cooler than  compact grains.
At the same time, the temperature of very porous grains becomes slightly larger
in the case of the EMT-Mie calculations in comparison with the
results found from the layered-sphere model.
The layered-sphere model predicts a broadening of infrared bands and
a shift of the peak position to larger wavelengths as  porosity
grows. In the case of the EMT-Mie model variations of the feature profile
are less significant.
It is also shown that the millimeter mass absorption coefficients
grow  as porosity increases with a faster growth occurring
for particles with Rayleigh/non-Rayleigh inclusions. As a result,
for very porous particles the coefficients given
by two models can differ by a factor of about 3.
\keywords{Scattering -- dust, extinction -- comets -- interplanetary medium --
      Stars: individual: $\zeta$ Oph,  $\sigma$ Sco --
      Galaxy: center}
      }
\maketitle

\section{Introduction}

The steady decrease of estimates of metal abundances in
the solar atmosphere  over the
last years (Holweger \cite{hol01}, Lodders \cite{lod03}, Asplund et al. \cite{ags04})
is  a serious {\rm challenge} not only {\rm to} solar physics, but
also to dust modelling.
 This calls for new {\rm dust} models  able to produce the same extinction
with a smaller amount of solid material.
A solution to the problem could be provided by
an ``admixture of vacuum'' i.e. by the porosity of interstellar grains.

Grain aggregates with {\rm large} voids can {\rm form} during
the growth of interstellar grains due to their coagulation
in dense molecular cloud cores (Dorschner \& Henning \cite{dh95}).
 The internal structure of such composite grains can
be very complicated, but
their optical properties are {\rm often} calculated using the Mie theory
for homogeneous spheres with an average refractive index
derived from effective medium theory
(EMT; see, e.g., Mathis \&  Whiffen \cite{mw89},
Jones \cite{jones88}, Ossenkopf \cite{oss91},
Mathis \cite{m96}, Li \& Greenberg \cite{li:gre98}, Il'in \& Krivova \cite{ik00}).

 Another approach  to calculate the optical properties of
such aggregates is the application of complex, computationally time consuming
methods such as the discrete dipole approximation (DDA;
see, e.g., Wright \cite{wr87}, Kozasa et al.~\cite{kbm92}, \cite{ko93},
Wolff et al.~{\cite{wo94}}, Stognienko et al.~\cite{st95},
Kimura \& Mann \cite{ki98}).

 Using the DDA,  Voshchinnikov et al.~(\cite{vih05}) examined
the ability of the EMT-Mie approach to treat porous
particles of different structure.
They show that the latter approach  can give relatively
accurate results only if the very porous particles have small
(in comparison with the wavelength of incident radiation) ``Rayleigh''
inclusions.
Otherwise, the approach becomes inaccurate
when the porosity exceeds $\sim$0.5.
 At the same time,  the optical properties of heterogeneous
spherical particles having inclusions of various sizes
(Rayleigh and non-Rayleigh) and very large porosity
were found to closely resemble  those of
spheres with a large number ($\ga 15-20$) of different layers.
The errors in extinction efficiency factors are smaller than 10~--~20\% if
the size parameter $\la 15$ and porosity is equal to $0.9$.
Note that this consideration was restricted by spheres,
not very absorbing materials (silicate and amorphous carbon)
and the integral scattering characteristics
(extinction, scattering, absorption efficiency factors,  albedo
and asymmetry parameter)
but not the differential cross sections or elements of the
scattering matrix.
Nevertheless, very simple computational models instead of
time-consuming DDA calculations give us a useful way to treat
composite grains of different structure.

In this paper,  we apply the
particle models of porous interstellar dust grains
based on the EMT-Mie and layered-sphere calculations.
The models described
in Sect.~\ref{model} are assumed to represent composite
particles with small (Rayleigh) inclusions and inclusions
of different sizes (Rayleigh and non-Rayleigh).
 The wavelength dependence of extinction is discussed in Sect.~\ref{ext1}.
 Sections~\ref{st_ext} and \ref{ir_ext}
 contain the application of the models to
calculations of the extinction curves in the directions
of two stars,  using new solar abundances
and the near-infrared (IR) extinction in the directions along the
Galactic plane.
 The next sections deal with grain temperatures
(Sect.~\ref{temp}), profiles of IR silicate bands (Sect.~\ref{ir_b}),
and grain opacities at $\lambda = 1$\,~mm (Sect.~\ref{opa}).
These quantities are especially important for the analysis of observations
of protoplanetary discs (Henning et al. \cite{heal05}).
Concluding remarks are presented in Sect.~\ref{concl}.

\section{Particle models}\label{model}

Information about the structure of grains can be included in light
scattering calculations directly (layered spheres) or can be
used to find the optical constants (EMT-Mie model).
We consider models of both types.

Following  previous papers (Voshchinnikov \&  Mathis \cite{vm99},
Voshchinnikov et al. \cite{vih05}),
we construct layered grains as particles consisting of many concentric
spherical layers of various materials, each with a specified
volume fraction $V_i$ ($\Sigma_i V_i /V_{\rm total} = 1)$.
Vacuum can be one of the materials, so a
composite particle may have a central cavity or voids in the form
of concentric layers.
The number of layers is taken to be 18 {\rm since}
Voshchinnikov et al.~(\cite{vih05})  have shown that this was enough
to preclude an influence of the order of materials on the results.
For a larger number of layers,
one can speak of the optical characteristics determined  by
the volume fractions of different constituents only.

In the case of the EMT-Mie model, an average (effective) refractive index
is calculated using  the popular rule of Bruggeman
(see, e.g., Ch\'ylek et al. \cite{cval00}, Kr\"ugel \cite{kru03}).
In this case, the  average dielectric permittivity
$\ve_{\rm eff}$$^($\footnote{The dielectric permittivity
is related to the refractive index as $\ve=m^2$.}$^)$
is calculated from
\be
\sum_i f_{i} \frac{\ve_{i} - \ve_{\rm eff}}{\ve_{i} + 2 \ve_{\rm eff}} = 0,
\label{bru}
\ee
where $f_{i}=V_i /V_{\rm total}$ is the volume fraction
of the $i$th material with the permittivity $\ve_{i}$.

The amount of vacuum in a particle can be characterized by its porosity
${\cal P}$ ($0 \leq {\cal P} < 1$)
\be
{\cal P} = V_{\rm vac} /V_{\rm total}
= 1 - V_{\rm solid} /V_{\rm total}. \label{por}
\ee
To compare the optical properties of porous and compact particles,
one should consider the  porous particles of radius (or size parameter)
\be
r_{\rm porous} = \frac{r_{\rm compact}}{(1-{\cal P})^{1/3}}
= \frac{r_{\rm compact}}{(V_{\rm solid} /V_{\rm total})^{1/3}}. \label{xpor}
\ee

 As ``basic'' {\rm constituents}, we choose
amorphous carbon (AC1; Rouleau \& Martin \cite{rm91})
and astronomical silicate (astrosil; Laor \&  Draine \cite{laordr93}).
 The refractive indices for {\rm these materials and some others} considered
in Sect.~\ref{ab_ext} were taken from the Jena--Petersburg Database of
Optical Constants (JPDOC) described by Henning et al.~(\cite{heal99})
and J\"ager et al.~(\cite{jetal02}).

The application of the standard EMT is known to correspond to the case
of particles having small randomly located inclusions of
different materials (Bohren \& Huffman \cite{bh83}).
 The optical properties of such particles have been well studied
(see, e.g., Voshchinnikov \cite{v02} and references therein).

 However, one needs the DDA or other computationally
time consuming techniques to treat  particles with inclusions
larger  than the wavelength of incident radiation.
The difference in the optical characteristics of particles with
small and large inclusions has been discussed in
previous studies (e.g., Wolff et al.~\cite{wo94}, \cite{wo98}).
The  fact that this difference drastically
grows with the porosity ${\cal P}$ and becomes quite essential
already for $ {\cal P} \ga 0.5$ has been discovered
{\rm only recently} by Voshchinnikov et al.~(\cite{vih05}).
 They also found that the scattering properties of
particles with inclusions of different sizes (including
those with sizes comparable to the wavelength) were
very close to those of layered
particles, having the same size and material volume fractions.
A similar conclusion was reached for an ensemble of  particles
where each particle has inclusions of one (but different) size only.
In both cases of ``internal'' or ``external'' mixing, the model
of layered spheres can be applied.

These results are of particular importance for applications including the
astronomical ones where we deal with very porous particles.
Note that instead of time consuming calculations with DDA-like
codes,  one can use the exact solution to the light scattering
problem for layered spheres, which is as quick as Mie theory, to estimate
the properties of particles with Rayleigh/non-Rayleigh inclusions.

Thus, the models of the homogeneous sphere (with EMT) and layered spheres,
both having fast implementations, allow one to probe the difference
in optical properties caused by the different structure of
scatterers.

\section{Interstellar extinction and interstellar abundances}\label{ab_ext}

\subsection{Wavelength dependence of extinction}\label{ext1}

\begin{figure}[htb]
\bc
\resizebox{\hsize}{!}{\includegraphics{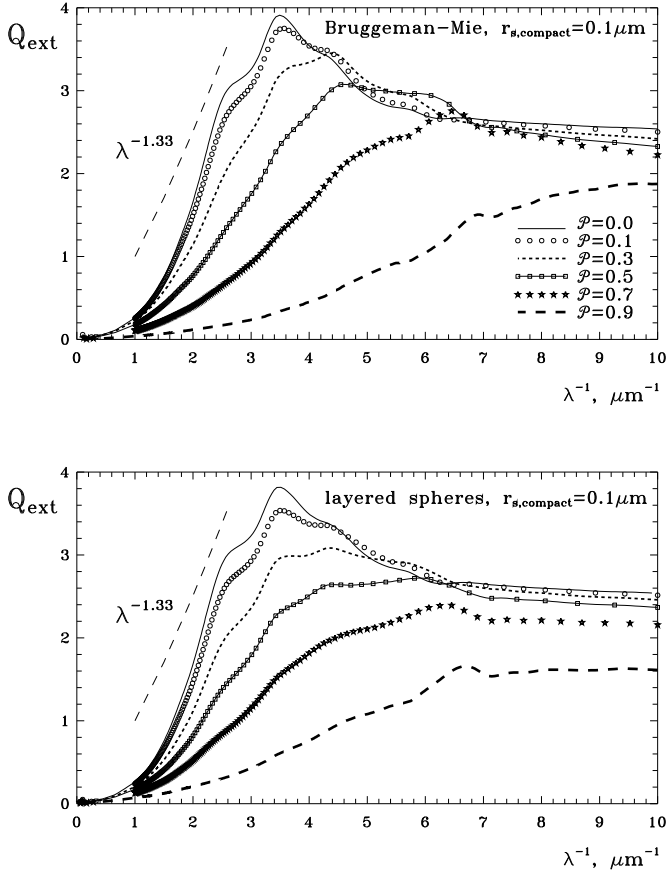}}
\caption{Wavelength dependence of the extinction efficiency {\rm factor}
for spherical particles with $r_{\rm s,\,compact}=0.1\,\mkm$.
The particles are of the same mass but of different porosity.
Upper panel: calculations based on the EMT-Mie theory.
Lower panel: calculations based on the layered-sphere theory.
}\label{ext_w}\ec
\end{figure}

As it is well known,
the wavelength dependence of  interstellar extinction $A(\lambda)$
is determined by the wavelength dependence of the
extinction efficiencies $Q_{\rm ext}(\lambda)$.
 This quantity is shown in Fig.~\ref{ext_w} for particles of the same mass,
but different porosity.
The volume fractions of AC1 and astrosil are equal, i.e.
$V_{\rm AC1} /V_{\rm total} = V_{\rm astrosil} /V_{\rm total} =
1/2 \, (V_{\rm solid} /V_{\rm total}) = 1/2 \, (1 - {\cal P})$.
 The radius of compact grains is  $r_{\rm s,\,compact}=0.1\,\mkm$.
 The dependence of $Q_{\rm ext}$ on $\lambda$ for compact particles
is close to that of the average interstellar extinction curve in the
visible--near UV ($1 \,\mkm^{-1} \leq\lambda^{-1} \leq 3 \,\mkm^{-1}$)
{\rm where it} can be approximated by the power law
$A(\lambda) \propto \lambda^{-1.33}$
(see discussion in Voshchinnikov~\cite{v02}).
 From Fig.~\ref{ext_w} one can conclude that the  extinction
produced by particles with small inclusions and layers
 differs   little  for compact
and slightly porous particles. The difference becomes {\rm most} pronounced in
the near and far--UV (for $\lambda^{-1} \ga 2.5 - 3 \,\mkm^{-1}$).

As follows from Fig.~\ref{ext_w}, the wavelength
dependence of extinction  flattens as porosity increases.
It is well known (see, e.g., Greenberg \cite{g78}) that
{\rm different particles produce comparable extinction if the products of their
size $r$ and refractive index are close, i.e.}
\be
 r \, |m-1| \approx {\rm const.}
\label{mr}\ee
 The average refractive index of particles with a larger fraction of vacuum
is closer to 1.
Despite a larger radius (e.g., from Eq.~(\ref{xpor}) follows that
$r_{\rm s}=0.22 \,\mkm$ if ${\cal P}=0.9$ and $r_{\rm s,\,compact}=0.1 \,\mkm$),
the product given by Eq.~(\ref{mr}) decreases because of {\rm an}
drop of $|m-1|$. This implies that a steeper extinction
with the wavelength dependence closer to $\propto \lambda^{-1.33}$
is produced by  for compact particles with larger radii and,
consequently, a larger amount of solid material.
This behaviour is {\rm also observed for} particles of other masses (i.e.,
compact spheres of other radii). Therefore, an interpretation
of the observed interstellar extinction curve
using only very porous grains should
not give any gain in dust-phase abundances and would contradict
the wavelength behaviour.

\begin{figure}\bc
\resizebox{\hsize}{!}{\includegraphics{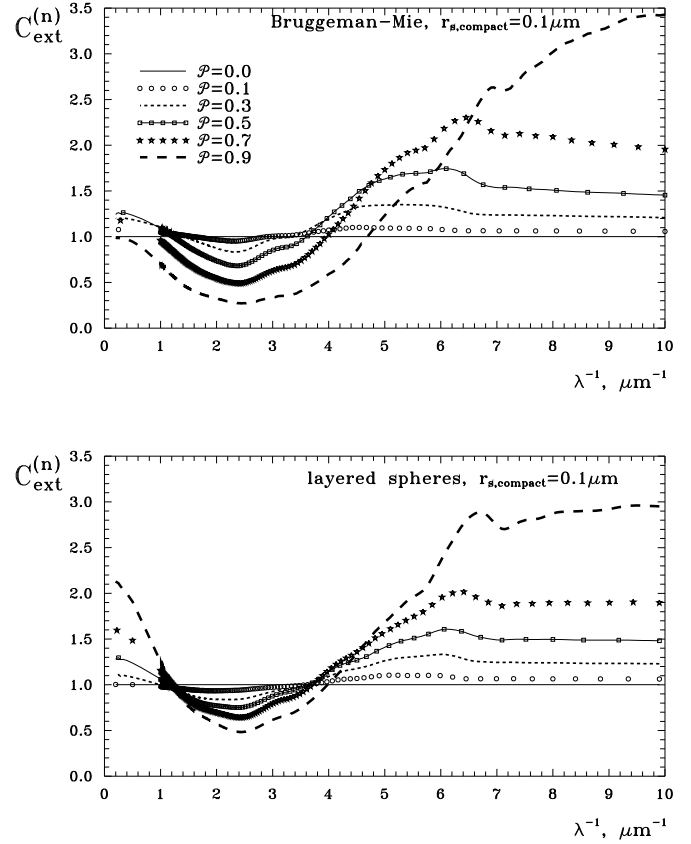}}
\caption{
The same as in Fig.~\ref{ext_w} but now for normalized extinction cross section.
}\label{cn_w}\ec
\end{figure}
\begin{figure}\bc
\resizebox{\hsize}{!}{\includegraphics{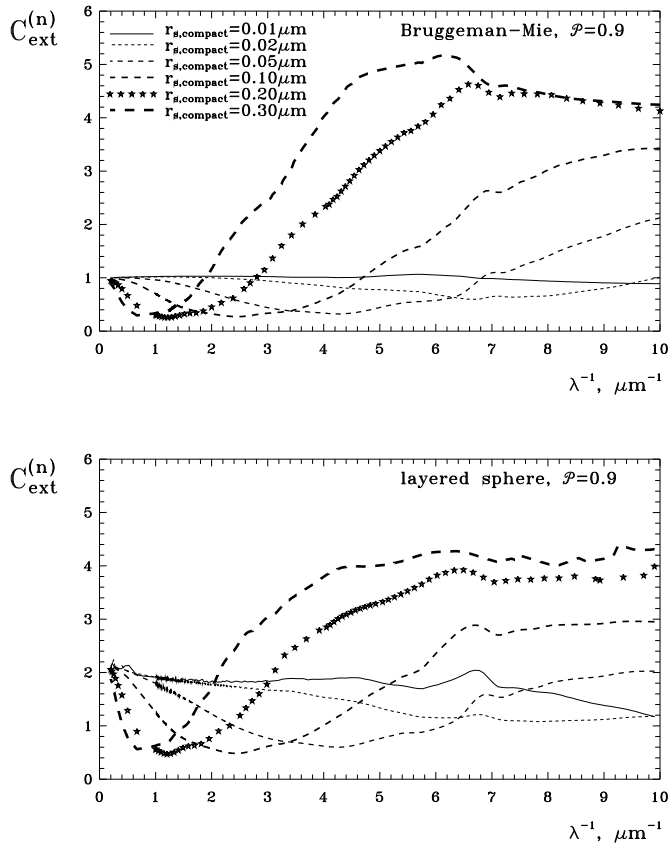}}
\caption{Wavelength dependence of the normalized extinction cross section
for spherical particles with the same  porosity ${\cal P}=0.9$.
Upper panel: calculations based on the EMT-Mie theory.
Lower panel: calculations based on the layered-sphere theory.
}\label{pp09}\ec
\end{figure}

The role of porosity in extinction is better seen from
Fig.~\ref{cn_w} where
{\rm we give} the wavelength dependence of the normalized cross section
\bea
C_{\rm ext}^{\rm (n)} = \frac{C_{\rm ext}({\rm porous \, grain})}
{C_{\rm ext}({\rm  compact \, grain \, of \, same \, mass})} = \nonumber \\
\,\,\,\,\,\,\, (1-{\cal P})^{-2/3}\,  \frac{Q_{\rm ext}({\rm porous \, grain})}
{Q_{\rm ext}({\rm  compact \, grain \, of \, same \, mass})}. \label{cn}
\eea
This quantity shows how porosity influences the extinction cross section.
As follows from Fig.~\ref{cn_w}, both models predict a growth of
extinction of porous particles in the far-UV and a decrease in the
visual--near-UV part  with growing ${\cal P}$. However, the wavelength interval
where $C_{\rm ext}^{\rm (n)} <1$ is  narrower and the minimum is
less deep in the case of layered spheres.
In comparison with compact grains and particles with Rayleigh inclusions,
particles with Rayleigh/non-Rayleigh inclusions can also produce rather
large extinction in the near-IR part of spectrum.
This is especially
important for the explanation of the flat extinction across the $3 - 8\,\mkm$
wavelength range measured for several lines of sight
(see Sect.~\ref{ir_ext}).
At the same time, as follows from Fig.~\ref{pp09},
for particles with $r_{\rm s,\,compact}>0.1 \,\mkm$
the normalized UV cross sections grow faster for the
Bruggeman--Mie theory than for layered spheres.
Thus, an addition of vacuum into particles does not mean an automatic
growth of extinction at all wavelengths and a saving in terms
of solid-phase elements.
Evidently, the final decision can be made after fitting the theoretical
calculations with observations at many wavelengths.

\subsection{Extinction in the directions
to the stars $\zeta$ Oph and $\sigma$ Sco}\label{st_ext}

The basic requirement for any model of interstellar dust is
the explanation of the observed extinction law along with
the dust-phase abundances of elements in the interstellar medium.
These abundances are obtained as the difference between the
cosmic reference abundances and the observed gas-phase abundances.
However, the cosmic abundances are not yet
conclusively established and usually this causes a problem.
For many years, the solar abundances were used as the reference ones,
until the photospheres of the early-type stars were found
not to be as rich in heavy elements as the solar
photosphere was (Snow \& Witt \cite{sw96}).
These stellar abundances caused the so-called ``carbon crisis''.
Abundances of the most important dust-forming elements (C, O, Mg, Si, Fe)
required by the  dust models were larger than available.
However, during the past several years the solar abundances
dropped and now they approach the stellar ones
(see Asplund et al. \cite{ags04}).
Evidently, some abundances determined by Sofia \& Meyer (\cite{sm01})
for F and G stars must be revised downward
as it has been done recently
for the Sun. This should lead to the agreement between abundances found
for stars of different types.
Note also that the current solar abundances of oxygen and iron
(see Table~\ref{ida}) are close to those found from
high-resolution \mbox{X-ray} spectroscopy.
Juett et al.~(\cite{jsc03}) investigated the oxygen
K-shell interstellar absorption edge in seven X-ray binaries and evaluated
the total O abundances. These abundances lie between
467~ppm$^{(}$\footnote{parts per million hydrogen atoms}$^{)}$ and 492~ppm.
Schulz et al.~(\cite{schet02}) evaluated the
total abundance of iron towards the object Cyg~X-1 to be
$[{\rm Fe}/{\rm H}]_{\rm cosmic} \approx 25$~ppm.

We applied the model of multi-layered porous particles
to explain the absolute extinction in the direction to the two stars.
A first estimate has been made in order to find a possibility
to enlarge the extinction per unit  mass and to minimize the amount
of solid phase material.
Several materials as components of composite grains were considered.
Among the carbonaceous species,
the amorphous carbon in the form of Be1 (Rouleau \& Martin \cite{rm91})
was found to produce the largest extinction. Also the extinction of iron
oxides strongly increases with the growth of porosity.
Although there are no very good constraints on the abundance of oxides,
FeO is considered as a possible carrier of the 21~$\mkm$ emission
observed in the spectra of post-AGB stars (Posch et al. \cite{pma04}).
Very likely, such particles particles can be produced in redox
reactions (Duley \cite{dul80}, Jones \cite{jones90}).

\begin{figure}[htb]
\begin{center}
\resizebox{\hsize}{!}{\includegraphics{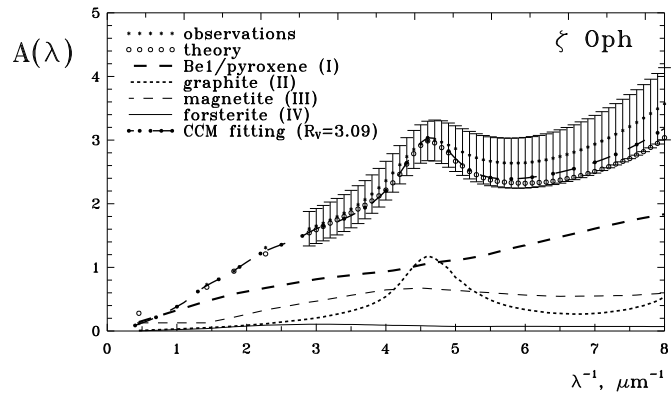}}
\caption
{Observed and calculated extinction in the direction to
$\zeta$ Oph.
The errors of the observations are the result of a parameterization
of the observations (see Fitzpatrick \&  Massa \cite{fm90}).
The contribution to the theoretical extinction from different components
is also shown.
{\rm The dot-dashed curve is the approximation with the observed value
$R_{\rm V}$ as suggested by Cardelli et al. (\cite{ccm89}).}
}
\label{zeta}\end{center}\end{figure}

\begin{figure}[htb]
\begin{center}
\resizebox{\hsize}{!}{\includegraphics{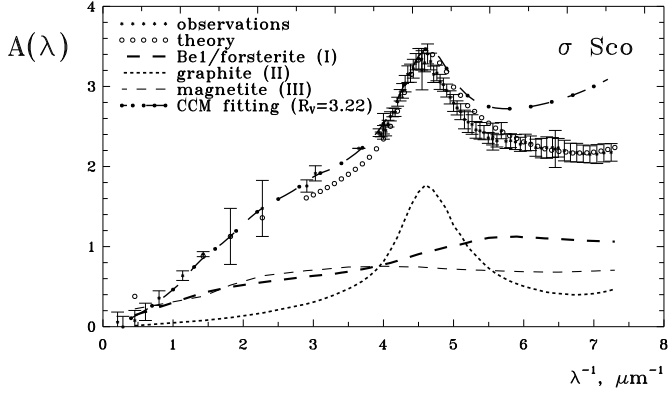}}
\caption
{The same as in Fig.~\ref{zeta} but now for  $\sigma$ Sco.
The observational data were taken from  Wegner~(\cite{ww02}).
}\label{sigma}\end{center}\end{figure}

We fitted the observed extinction  toward the stars
$\zeta$~Oph  and $\sigma$ Sco.
For these stars there exist well determined extinction curves and
gas-phase abundances. It is also important that the major
part of extinction in these directions is produced in one
diffuse interstellar cloud (Savage \&  Sembach \cite{ss96},
Zubko et al. \cite{zkw96}). This allows us to exclude possible large
variations in dust composition along the line of sight.

Observed and calculated extinction curves are plotted in
Figs.~\ref{zeta} ($\zeta$~Oph)  and \ref{sigma} ($\sigma$ Sco).
As  follows from the previous Section,
the use of only porous or only compact grains apparently
does not result significant benefit in the   solid-phase abundances.
Therefore, our models are the combination of compact and
porous particles. They consist of three or four grain populations:

(I). Porous composite (multi-layered) particles
(Be1 --- 5\%,
pyroxene, Fe$_{0.5}$Mg$_{0.5}$SiO$_3$ --- 5\% for $\zeta$ Oph or
forsterite, Mg$_2$SiO$_4$ --- 5\% for $\sigma$ Sco and
vacuum --- 90\%)
with the power-law size distribution having an exponential decay
$n_{\rm d}(r_{\rm s})\propto r_{\rm s}^{-2.5}\exp (-10/r_{\rm s})$.
The lower/upper cut-off in the size distribution is
$0.015\,\mkm$/$0.25\,\mkm$ and $0.05\,\mkm$/$0.50\,\mkm$
for $\zeta$~Oph and $\sigma$ Sco, respectively.

(II). Small compact graphite\footnote{
The calculations for graphite were made in the ``2/3--1/3'' approximation:
$Q_{\rm ext} = {2}/{3}\, Q_{\rm ext}(\varepsilon_{\bot}) +
              {1}/{3}\, Q_{\rm ext}(\varepsilon_{||})$,
where $\varepsilon_{\bot}$ and $\varepsilon_{||}$ are the dielectric
functions for two cases of orientation of the electric field relative
to the basal plane of graphite.} grains
with a narrow power-law size distribution
($n_{\rm d}(r_{\rm s})\propto r_{\rm s}^{-2.5}$,
$r_{\rm s}=0.01 - 0.02\,\mkm$).

(III). Porous composite grains of magnetite
(Fe$_3$O$_4$ --- 2\%, vacuum --- 98\% for $\zeta$ Oph and
Fe$_3$O$_4$ --- 8\%, vacuum --- 92\%  for $\sigma$ Sco)
with a power-law size distribution
$n_{\rm d}(r_{\rm s})\propto r_{\rm s}^{-2.5}$.
The lower/upper cut-off in the size distribution is
$0.005\,\mkm$/$0.25\,\mkm$ and $0.05\,\mkm$/$0.35\,\mkm$
for $\zeta$~Oph and $\sigma$ Sco, respectively.

(IV). Compact grains of forsterite (Mg$_2$SiO$_4$)
with the power-law size distribution (only for $\zeta$ Oph,
$n_{\rm d}(r_{\rm s})\propto r_{\rm s}^{-3.5}$,
$r_{\rm s,\,min}=0.10\,\mkm$, $r_{\rm s,\,max}=0.25\,\mkm$). \\

\begin{table*}[htb]
\bc
\caption[]{Contribution of different grain populations
to $A_{\rm V}$ and dust-phase abundances
for the model of $\zeta$ Oph (in ppm)}\label{da-oph}
\begin{tabular}{llccccc}
\hline
\noalign{\smallskip}
Component & $A_{\rm V}$  & C & O & Mg& Si & Fe \\
\noalign{\smallskip}
\hline
\noalign{\smallskip}
(I)   Be1/pyroxene/vacuum & 0\fm58 & 123  &  68  & 11.3 & 22.6 & 11.3  \\
(II)  Graphite & 0\fm075&  ~96  &      &      &      &      \\
(III) Magnetite/vacuum & 0\fm22&      &  33  &      &      & 24.8 \\
(IV) Forsterite & 0\fm065&      &  23  & 11.4 &  5.7 &      \\
\noalign{\smallskip}\hline
\noalign{\smallskip}
Total & 0\fm94 & 219 & 124 & 22.7 & 28.2 & 36.1\\
\noalign{\smallskip}\hline
\end{tabular}\ec
\end{table*}

Figures~\ref{zeta} and \ref{sigma} also contain
the extinction curves calculated using the approximation
suggested by Cardelli et al. (\cite{ccm89}) with
the coefficients revised by O'Donnell (\cite{odonn94}).
Cardelli et al. (\cite{ccm89}) found that
the extinction curves from the UV through the IR
could be characterized as a one-parameter family dependent
on the ratio of the total extinction to the selective one
$R_{\rm V}=A_{\rm V}/E({\rm B-V})$.
We used the observed values of $R_{\rm V}$ in order
to plot the CCM approximation. It is seen that
this relation describes quite well the extinction for
$\zeta$~Oph  but not for $\sigma$ Sco.

\begin{table*}[htb]
\bc
\caption[]{Contribution of different grain populations
to $A_{\rm V}$ and dust-phase abundances
for the model of $\sigma$ Sco (in ppm)}\label{da-sco}
\begin{tabular}{llccccc}
\hline
\noalign{\smallskip}
Component & $A_{\rm V}$ & C & O & Mg& Si & Fe \\
\noalign{\smallskip}
\hline
\noalign{\smallskip}
(I)   Be1/forsterite/vacuum & 0\fm50 & 58  &  35.4 & 17.7 & 8.8 &   \\
(II)  Graphite              & 0\fm11&  79  &      &      &      &      \\
(III) Magnetite/vacuum      & 0\fm52&      &  35.4 &      &      & 26.6 \\
\noalign{\smallskip}\hline\noalign{\smallskip}
Total & 1\fm13 & 137 & 71 & 17.7 & 8.8 & 26.6\\
\noalign{\smallskip}\hline
\end{tabular}\ec
\end{table*}

The contributions from different components to the calculated extinction
are given in Tables~\ref{da-oph} and \ref{da-sco}
and shown in Figs.~\ref{zeta} and \ref{sigma}.
The Tables contain also the dust-phase abundances
of five dust-forming elements for several grain populations.
They were calculated for ratios of the extinction cross-section
to particle volume averaged over grain size distribution
(see Eq.~(3.36) in Voshchinnikov~\cite{v02}).

Table~\ref{ida} gives the current solar abundances
of five dust-forming elements according to Asplund et al.~(\cite{ags04})
as well as the ``observed''
(solar minus gaseous)  and model  abundances.

The dust-phase abundances in the line of sight to the star
$\zeta$~Oph (HD~149757) were taken from
Table~2 of Snow \& Witt~(\cite{sw96}).
In our calculations, we adopted
the following quantities
for $\zeta$~Oph: a total extinction
$A_{\rm V}=0\fm94^($\footnote{This value was obtained from the
relation $A_{\rm V}= 1.12 \, E({\rm V-K})$ (Voshchinnikov  \& Il'in \cite{vi87})
and a colour excess  $E({\rm V-K})=0\fm84$ (Serkowski et al. \cite{smf75}).}$^)$,
 colour excess $E({\rm B}-{\rm V})=0\fm32$ and
total hydrogen column density $N({\rm H})=1.35\,10^{21}\,{\rm cm}^{-2}$
(Savage \&  Sembach, \cite{ss96}).
The extinction curve was reproduced according to the parameterization
of Fitzpatrick \&  Massa~(\cite{fm90}).

For $\sigma$ Sco (HD~147165), we used the extinction curve,
the colour excess $E({\rm B}-{\rm V})=0\fm35$ and
the total extinction $A_{\rm V}=1\fm13$ according to Wegner~(\cite{ww02}).
The hydrogen column density
$N({\rm H})=2.46 \, 10^{21}\, {\rm cm}^{-2}$ was adopted from
Zubko et al.~(\cite{zkw96}). The gas-phase abundances were taken from
Allen et al.~(\cite{asj90}).

The dust-phase abundances required by the model  are larger
than the observed ones in the direction to $\zeta$~Oph
(for C and Fe) and
smaller than the observed abundances in the direction to $\sigma$ Sco.
Note that for $\sigma$ Sco the required amount of C and Si in dust
grains is the lowest in  comparison with  previous
modelling. This is due to the use of highly porous particles
which give  considerable extinction in the UV and near-IR (see
Figs.~\ref{cn_w} and \ref{pp09}) and  allow one to ``save''  material.
For example, the extinction model of $\sigma$~Sco with compact grains
presented by Zubko et al.~(\cite{zkw96})
requires 240~--~260 ppm of C and 20~--~30 ppm of Si
and the model of Clayton et al.~(\cite{cletal03})
needs 155 ppm of C and 36 ppm of Si
(cf. 137 ppm and 8.8 ppm from Table~\ref{ida}).

The models presented above are based on the light scattering
calculations for particles with  Rayleigh/non-Rayleigh inclusions
(layered spheres). It is evident that the observed extinction
can be also reproduced if we use the particles with  Rayleigh inclusions
(i.e., if we apply the EMT-Mie theory). Our estimates show that
despite a larger extinction in the UV
this model requires more material in the solid phase
in comparison with layered spheres because of
a smaller extinction in the visual-near IR part of the spectrum.

\begin{table}[htb]
\bc
\caption[]{Observed and model dust-phase  abundances (in ppm)}\label{ida}
\begin{tabular}{cccccc}
\hline
\noalign{\smallskip}
Element & Solar$^\ast$  & \multicolumn{2}{c} {$\zeta$ Oph} &\multicolumn{2}{c} {$\sigma$ Sco}\\
 & abundance& obs & model & obs & model \\
\noalign{\smallskip}
\hline
\noalign{\smallskip}
~C&              245  & 110  &  219  &     176    & 137    \\
~O&              457  & 126  &  124  &     ~85    & ~71    \\
Mg$^{\ast\ast}$& ~33.9 & ~31.9&  ~22.7&   ~30.9    & ~~~17.7 \\
Si&             ~34.2 & ~32.6&  ~28.2&  ~~32.4    & ~~~8.8  \\
Fe&             ~28.2 & ~28.2&  ~36.1& ~~~27.9    & ~~~26.6  \\
\noalign{\smallskip}\hline
\end{tabular}\ec
\noi $^\ast$ According to Asplund et al.~(\cite{ags04}). \\
\noi $^{\ast\ast}$ The abundance of Mg was recalculated with the  oscillator
strengths from Fitzpatrick~(\cite{fp97}).

\end{table}

\subsection{Near infrared extinction in the Galactic plane}\label{ir_ext}

We now consider the possibility  of explaining the flat extinction
across the $3 - 8\,\mkm$ wavelength range observed for
several lines of sight. This flattening was first measured by
Lutz et al.~(\cite{luetal96}) toward the Galactic center
with {\it ISO}, using hydrogen recombination lines.
Later Lutz~(\cite{lutz99}) confirmed the effect using more
recombination lines. Recently, Indebetouw et al.~(\cite{inetal05})
found a similar flat extinction along two lines of sight:
$l=42\degr$ and $l=284\degr$. The extinction was obtained at seven
wavelengths ($1.2 - 8\,\mkm$) by combining images from the {\it Spitzer
Space Telescope} with the 2MASS point-source catalog.

\begin{figure}[htb]
\begin{center}
\resizebox{\hsize}{!}{\includegraphics{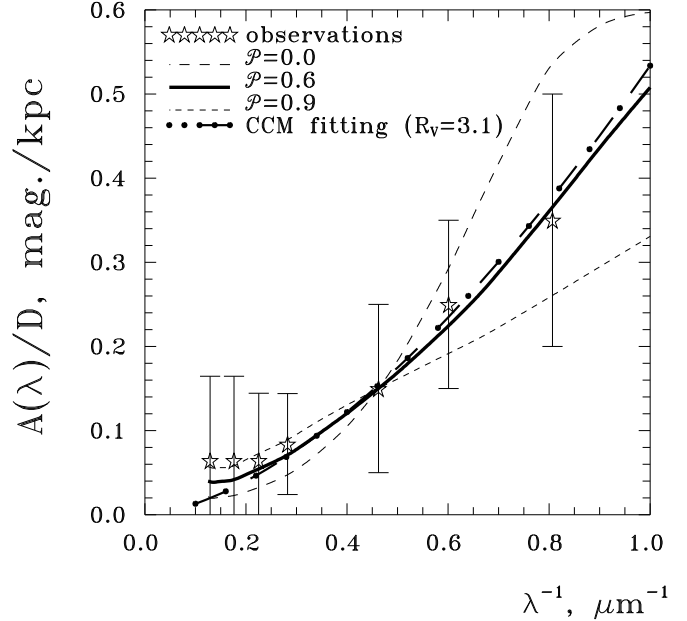}}
\caption
{Observed and calculated extinction in the near-IR part
of spectrum. The observations correspond to the average extinction
for two lines of sight along the Galactic plane
(Indebetouw et al.~\cite{inetal05})
transformed into magnitudes of extinction per kpc.
The theoretical extinction was calculated for component~(I)
of the model used for  $\zeta$ Oph (${\cal P}=0.9$, short dashed curve).
Two other curves correspond to the same component but with
another particle porosity.
{\rm The dot-dashed curve is the approximation of Cardelli et al. (\cite{ccm89})
with $R_{\rm V}=3.1$.}
}
\label{gc}\end{center}\end{figure}
We used the average extinction given in Table~1 of
Indebetouw et al.~(\cite{inetal05}) and transformed
it into magnitudes of extinction per kpc using
the measured value  $A_{\rm K}/D = 0\fm15 \pm 0\fm1\,{\rm kpc}^{-1}$
(Indebetouw et al.~\cite{inetal05}).
The observations are plotted in Fig.~\ref{gc} together with
three theoretical curves.
Because we have  little information about the UV-visual extinction and
gas-phase abundances in these directions, we (rather arbitrarily)
applied the model used for
$\zeta$ Oph (porous component (I): Be1/pyroxene, porosity 90\%; see
Sect.~\ref{st_ext}). This model (short dashed curve on Fig.~\ref{gc})
well explains the flat extinction at $\lambda > 3\,\mkm$$^($\footnote{Note
that porous particles from magnetite (component (III); see
Sect.~\ref{st_ext}) cannot fit well the extinction at these wavelengths
because of a bump at $\lambda \approx 2\,\mkm$.}$^)$
but the extinction in the J and H  bands is  too small.
Compact particles (long-dashed curve in Fig.~\ref{gc}) produce
even larger extinction at these bands than the observed one.
However, the extinction from such particles at longer wavelengths
decreases rapidly. Our preliminary analysis shows that
particles with a porosity of
about 0.6 (solid curve on Fig.~\ref{gc})
can be chosen as an appropriate model.
Evidently, a similar curve can be obtained as a combination
of compact and very porous particles.
Quite close extinction can be found with
the CCM approximation and the standard value $R_{\rm V}=3.1$.
We arbitrarily extrapolated this approximation to long wavelengths
($\lambda^{-1} < 0.3 \,\mkm^{-1}$) where it gives too small extinction.

 Extinction produced by porous grains was also rather flat
between 1.0 and 2.2~$\mkm$ (for example,
$A(\lambda) \propto \lambda^{-1.3}$ for ${\cal P}=0.6$) as was detected
for several ultracompact HII regions with $A_{\rm V} \ga 15^{\rm m}$
(Moore et al. \cite{metal05}).

In order to calculate the dust-phase  abundances
for models presented in Fig.~\ref{gc}
we estimated the total hydrogen column density
using Eqs.~(3.26), (3.27) and (3.22) from Voshchinnikov~(\cite{v02}).
First, we found  the column density of
atomic hydrogen $N({\rm HI})$ from the extinction at J band and then
transformed $N({\rm HI})$ into a total hydrogen column density
$N({\rm H})$ using the ratio of total to selective extinction
$R_{\rm V}=3.1$. A value of
$N({\rm H})/D=2.79 \, 10^{21}\, {\rm cm}^{-2}{\rm kpc}^{-1}$
was obtained.
The calculated abundances are given in Table~\ref{igc}, which also
contains the visual extinction calculated for three
models. Note that the model of grains with porosity ${\cal P}=0.6$
gives the largest contribution to $A_{\rm V}$ in comparison with
two other models.

\begin{table}[htb]
\bc
\caption[]{Dust-phase  abundances (in ppm)
and visual extinction for three models presented
in Fig.~\ref{gc}.}\label{igc}
\begin{tabular}{cccc}
\hline
\noalign{\smallskip}
Element & ${\cal P} = 0$ & ${\cal P} = 0.6$ & ${\cal P} = 0.9$\\
\noalign{\smallskip}
\hline
\noalign{\smallskip}
~C&      96  &  84  &   62   \\
~O&      53  &  46  &   34   \\
Mg&      ~8.9& ~7.7 &  ~5.7   \\
Si&      18  &  16  &   11    \\
Fe&    ~8.9  & ~7.7 &   ~5.7  \\
\noalign{\smallskip}\hline\noalign{\smallskip}
$A_{\rm V}$ &   0\fm60  & 0\fm79 &  0\fm61  \\
\noalign{\smallskip}\hline
\end{tabular}\ec
\end{table}

There are useful observational data
for foreground stars in the field $l=284\degr$.
For stars HD~90273,  HD~93205 and HD~93222,
Wegner~(\cite{ww02}) and Barbaro et al.~(\cite{betal04})
estimate  the values of $R_{\rm V}$ which lie between 3.4 and 4.0.
For HD~90273 Barbaro et al.~(\cite{betal04}) also find
an anomalously high gas to dust ratio
$N({\rm H})/E({\rm B}-{\rm V})=
1.04\,10^{22}\,{\rm atoms}\,{\rm cm}^{-2}\,{\rm mag.}^{-1}$
Enlargement of $R_{\rm V}$ increases the required dust-phase
abundances while the decrease of the gas to dust ratio  reduces them.
Andr\'e et al.~(\cite{and03}) measured the  interstellar gas-phase
oxygen abundances along the sight lines toward 5 early-type stars
with  $l=285\fdg3-287\fdg7$ and  $b=-5\fdg5~-~+0\fdg1$.
The values of $[{\rm O}/{\rm H}]_{\rm g} = 443$~ppm vary from
 356~ppm to 512~ppm, the average value being 443~ppm.
This gives for the mean dust-phase
abundance $[{\rm O}/{\rm H}]_{\rm d} = 14$~ppm.
However, the extinction curves for HD~93205 and HD~93222
published by Wegner~(\cite{ww02}) have strong UV bumps and flat
extinction in the far-UV. This means that extinction can be mainly produced
by carbonaceous grains.
Evidently, the most reasonable way to solve the problem of
abundances  is a re-examination of the reference cosmic
abnudances and a detailed study of their local values.

\section{Infrared radiation}\label{irr}
\subsection{Dust temperature}\label{temp}

\begin{figure}\bc
\resizebox{\hsize}{!}{\includegraphics{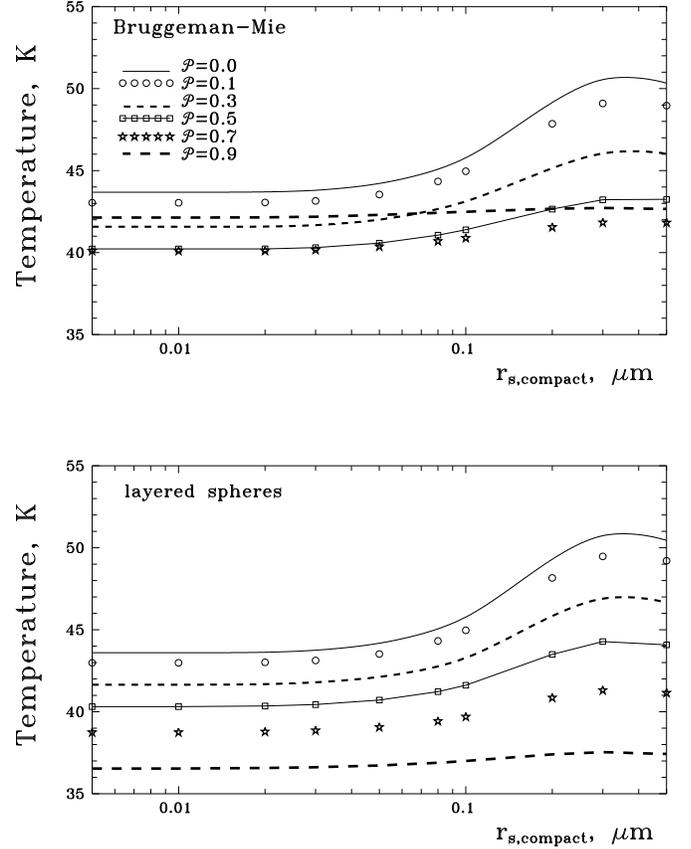}}
\caption{Size dependence  of the temperature
for spherical particles.
The particles are located at  a distance of
10$^4\,R_\star$ from a star with an effective temperature
$T_\star=2500$\,K.
Upper panel: calculations based on the EMT-Mie theory.
Lower panel: calculations based on the layered-sphere theory.
}\label{td}\ec
\end{figure}
\begin{figure}[htb]\bc
\resizebox{\hsize}{!}{\includegraphics{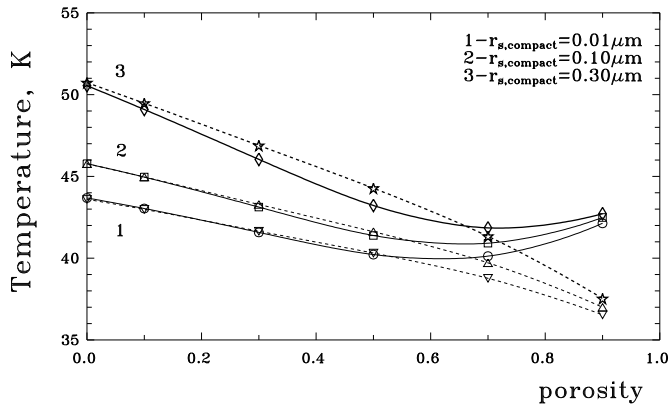}}
\caption{Dependence  of the dust temperature on particle porosity.
Solid lines: calculations based on the EMT-Mie theory.
dashed lines: calculations based on the layered-sphere theory.
Other parameters are the same as in Fig.~\ref{td}.
}\label{td-ppp}\ec\end{figure}

The commonly used equilibrium temperature
of cosmic grains is derived from a balance {\rm between} the energy gain
due to absorption of the UV and visual stellar photons
and  the energy loss due to re-emission of IR photons.
The temperature of porous and compact particles
of different size and porosity is shown in
Figs.~\ref{td} and \ref{td-ppp} as a function of particle
size and porosity, respectively.
The results were calculated for particles located at a distance of
10$^4\,R_\star$ from a star with an effective temperature
$T_\star=2500$\,K. In the case of the layered spheres,
an increase of the vacuum fraction
causes a decrease of the grain temperature if the  amount of the
solid material is kept constant. This behaviour holds for particles of all
sizes as well as for particles located closer to the star or farther away
and for other values of $T_\star$. If the  EMT-Mie theory is applied,
the temperature  drops when the porosity grows up to $\sim 0.7$ and then
starts to increase (see Fig.~\ref{td}, upper panel and Fig.~\ref{td-ppp}).
Such a behaviour corresponds to the results
of Greenberg \& Hage~(\cite{grha91}) who found an increase of
temperature for large grain porosity (see Fig.~4 in their paper).

As it follows from Figs.~\ref{td} and \ref{td-ppp},
the difference in the temperature of very porous grains
calculated using the two models can reach $\sim 6$\,K or  $\sim 15$\%
while the temperature of compact (${\cal P} = 0$)
composite grains differs by less than 1\%.
Note that the relative difference in temperatures of $\sim 15$\%
between particles of the two types is kept for other stellar temperatures
(e.g., in the case of the Sun or an interstellar radiation field).

The intermediate porosity of grains leads to a shift of the
peak position of IR emission to larger wavelengths
in comparison with compact particles. This occurs independently of
particle structure (small or various size inclusions).
However, very porous grains with Rayleigh/non-Rayleigh inclusions
are expected to be systematically cooler than
particles  with Rayleigh inclusions. Such a difference can be of
great importance at a lower temperature regime because it can influence
the growth/destruction of mantles on grains in molecular clouds.

\subsection{Infrared features}\label{ir_b}

It is well known that the shape of the IR dust features is a good indicator
of  the particle size and chemical composition.  With an increase of the size,
a  feature becomes wider and eventually fades away.
For example, in the case of compact spherical grains of astrosil,
the 10~$\mkm$ and  18~$\mkm$  features disappear when the grain radius
exceeds $\sim 2-3 \,\mkm$.
Observed differences in small scale structure of the features
are usually attributed to variations of  the composition
(e.g., changes of the ratio of magnesium
to iron in silicates) or  material state (amorphous/crystalline).

\begin{figure}\bc
\resizebox{\hsize}{!}{\includegraphics{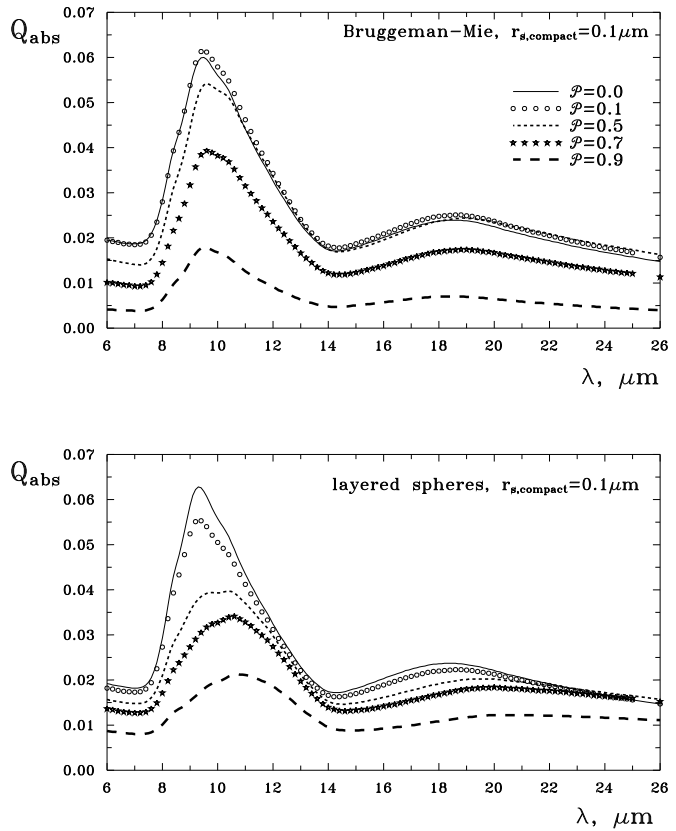}}
\caption{Wavelength dependence  of the absorption efficiency factors
for spherical particles
of radius $r_{\rm s, \, compact}=0.1 \,\mkm$.
Upper panel: calculations based on the EMT-Mie theory.
Lower panel: calculations based on the layered-sphere theory.
}\label{abs01}\ec
\end{figure}
In Fig.~\ref{abs01} we compare the wavelength dependence of
the absorption efficiency factors for particles of the same mass but different
structure.
 The upper panel shows  results obtained with the EMT-Mie model
for particles with Rayleigh inclusions.
It can be seen that the central position and the width of the dust features
does not really change.
Larger changes occur for the layered-sphere model
(Fig.~\ref{abs01}, lower panel).
 In this case a growth of ${\cal P}$ causes a shift of the center
of the feature to longer wavelengths and its broadening.
 For particles with ${\cal P}=0.9$, the 10~$\mkm$  feature transforms into
a plateau while the 18~$\mkm$  feature disappears.

\begin{figure}\bc
\resizebox{\hsize}{!}{\includegraphics{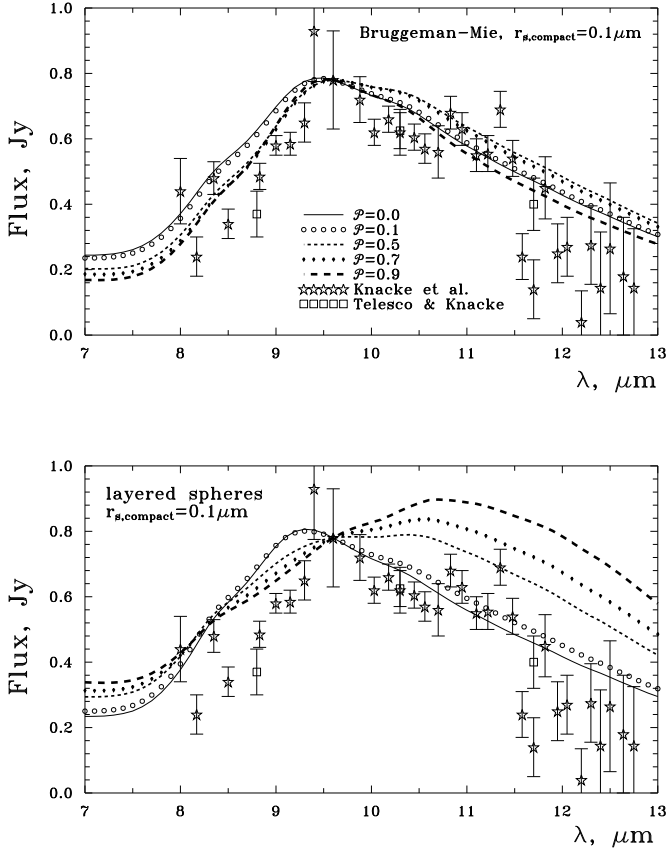}}
\caption[]
{Emission in the disc around the star $\bet$  in the region of silicate
10\,$\mkm$ band.
Stars and squares are the observations of Knacke et al.~(\cite{kn93})
and of Telesco \& Knacke~(\cite{tk91}). The curves present the results
of calculations for particles of radius $r_{\rm s, \, compact}=0.1 \,\mkm$
as shown in Fig.~\ref{abs01} but normalized at $\lambda=9.6\,\mkm$.
}\label{bet}\ec
\end{figure}
We plotted in Fig.~\ref{bet} our data from Fig.~\ref{abs01} in a normalized
manner together with observations of $\bet$ made by
Knacke et al.~(\cite{kn93}) and Telesco \& Knacke~(\cite{tk91}).
As follows from Fig.~\ref{bet}, for given optical constants of the silicate
the observed shape of the 10~$\mkm$ feature
is better reproduced  by either compact or porous particles
with small size inclusions of materials.
Note that in the case of  $\bet$ the EMT-Mie calculations
were earlier used by  Li \& Greenberg~(\cite{li:gre98}) for the
explanation of the 10~$\mkm$ emission feature and by
Voshchinnikov \& Kr\"ugel~(\cite{vk99}) for
the interpretation of the positional
and wavelength dependence of polarization.
The best fit was obtained for very
porous particles: ${\cal P} \approx 0.95$ and 0.76, respectively.

\begin{figure}\bc
\resizebox{\hsize}{!}{\includegraphics{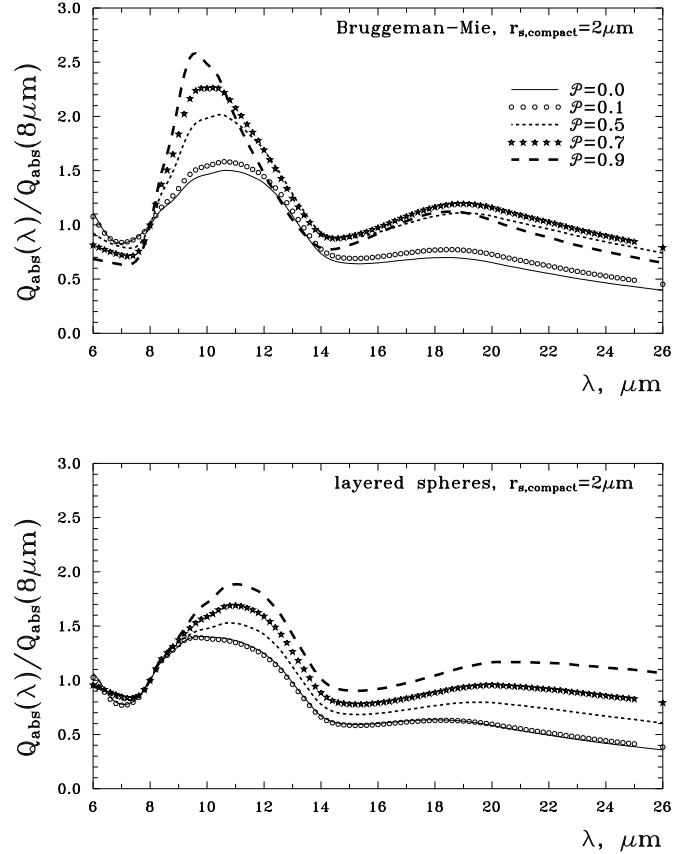}}
\caption{Wavelength dependence  of the normalized absorption efficiency factors
for spherical particles of radius $r_{\rm s, \, compact}=2 \,\mkm$.
Upper panel: calculations based on the EMT-Mie theory.
Lower panel: calculations based on the layered-sphere theory.
}\label{abs2}\ec
\end{figure}
Figure~\ref{abs2} shows  the normalized absorption efficiency factors
for spherical particles of radius $r_{\rm s, \, compact}=2 \,\mkm$.
For compact grains the 10~$\mkm$ feature almost disappears.
When the porosity increases the strength of the feature grows
in the case of the Bruggeman-Mie calculations.
This tendency coincides with the results
shown in Fig.~7 of Hage \& Greenberg~(\cite{hagr90}) who found
that the higher the porosity, the sharper the silicate emission {\rm became}.
For the case of layered spheres the feature becomes only slightly
stronger but its peak shifts to longer wavelengths.

A standard ``compact'' approach to the modelling of
the 10~$\mkm$  feature was used  by
van Boekel et al.~(\cite{vb03}, \cite{vb05})
and Przygodda et al. (\cite{pr03}) who considered the flattening of the 10~$\mkm$  feature
as an evidence of grain growth in the discs around Herbig Ae/Be
stars and T Tauri stars, respectively.
Our investigations show that the variations
of the shape of the feature and its position and strength can
also be attributed to the change of porosity and relative amount
of carbon in composite grains of small sizes.

\subsection{Dust opacities}\label{opa}

\begin{table*}[htb]
\caption[]{Mass absorption coefficients at $\lambda = 1$\,mm of
compact and porous spheres consisting of 
AC1$^\ast$ and (or) astrosil$^{\ast\ast}$.}    \label{t1}
\bc\begin{tabular}{cccccccccc}
\hline
\noalign{\smallskip}
&\multicolumn{3}{c}{AC1 + astrosil}&\multicolumn{3}{c}{astrosil}
&\multicolumn{3}{c}{AC1 } \\
\noalign{\smallskip}  \cline{2-10} \noalign{\smallskip}
${\cal P}$ &  $\rho_{\rm d}$&\multicolumn{2}{c}{$\kappa$, cm$^2$/g}
&  $\rho_{\rm d}$ &\multicolumn{2}{c}{$\kappa$, cm$^2$/g}
&  $\rho_{\rm d}$ &\multicolumn{2}{c}{$\kappa$, cm$^2$/g} \\
\noalign{\smallskip}  \cline{2-10} \noalign{\smallskip}
&&Brugg.--Mie & lay. spheres
&&Brugg.--Mie & lay. spheres
&&Brugg.--Mie & lay. spheres \\
\noalign{\smallskip}
\hline
\noalign{\smallskip}
 0.00 & ~2.58   &   1.58 &   1.58 & ~3.30   &   0.310&   0.310 & ~1.85  &   4.37 &   ~4.37\\
 0.10 & ~2.32   &   1.89 &   1.55 & ~2.97   &   0.371&   0.334 & ~1.66  &   5.13 &   ~4.60\\
 0.30 & ~1.80   &   2.75 &   1.87 & ~2.31   &   0.548&   0.446 & ~1.30  &   7.09 &   ~5.77\\
 0.50 & ~1.29   &   3.83 &   2.57 & ~1.65   &   0.778&   0.646 & 0.925  &   9.22 &   ~7.88\\
 0.70 & 0.772   &   3.94 &   4.04 & 0.990   &   0.794&  ~1.05  & 0.555  &   9.14 &  11.9 \\
 0.90 & 0.258   &   2.45 &   8.12 & 0.330   &   0.431&  ~2.20  & 0.185  &   5.94 &  21.8 \\
\noalign{\smallskip}
\hline
\noalign{\smallskip}
\end{tabular}\ec
 $^{\ast}$ $m(\lambda=1 {\rm mm})=2.93+0.276i$, $\rho_{\rm d}=1.85$\,g/cm$^3$

 $^{\ast\ast}$ $m(\lambda=1 {\rm mm})=3.43+0.050i$, $\rho_{\rm d}=3.3$\,g/cm$^3$
\end{table*}

The dust opacity or
the mass absorption coefficient of a grain material $\kappa(\lambda)$
enters directly in the expression for
the dust mass of an object $M_{\rm d}$ which is determined from
optically thin millimeter emission
\be
M_{\rm d} =  \frac{F_{\rm mm}(\lambda) D^2}{\kappa(\lambda) B_\lambda(T_{\rm d})}.
        \label{m}
\ee
Here, $F_{\rm mm}(\lambda)$ is  the observed flux,
$D$ the distance to the object, $B_\lambda(T_{\rm d})$ the Planck function,
$T_{\rm d}$ the dust temperature.
The mass absorption coefficient  $\kappa(\lambda)$ depends on
the particle volume $V_{\rm total}$,  the material density $\rho_{\rm d}$
and the extinction cross-section $C_{\rm ext}$ as follows:
\be
\kappa(\lambda) = \frac{C_{\rm ext}}{\rho_{\rm d} V_{\rm total}}  \approx
  \frac{3}{\rho_{\rm d}} \, \left(\frac{2 \pi}{\lambda}\right) \,
 {\rm Im} \left\{\frac{\ve_{\rm eff}-1}{\ve_{\rm eff}+2} \right\}.
        \label{kap}
\ee
At long wavelengths the scattering can be neglected
{\rm ($C_{\rm ext} \approx C_{\rm abs}$)} and
$C_{\rm abs}$ {\rm can be evaluated} in the Rayleigh approximation.
Then the mass absorption coefficient does not depend on the particle size
as shown in  the right part of Eq.~(\ref{kap}).
The effective dielectric permittivity $\ve_{\rm eff}$
in Eq.~(\ref{kap}) can be found from the Bruggeman rule (see Eq.~(\ref{bru})) or
the layered-sphere rule of the EMT (see Eqs.~(7), (8) in
Voshchinnikov et al. \cite{vih05}).

Extensive  studies of the mass absorption  coefficient dependence
on the material properties and grain shape are
summarized by Henning~(\cite{h96}) who,
in particular, {\rm notes} that the opacities at 1~mm are considerably
larger for non-spherical particles {\rm than for} spheres
(see also Ossenkopf \& Henning \cite{oh94}).
We find that a similar effect  (an increase of opacity in comparison
with compact spheres) is
produced by inclusion of a large fraction of vacuum
into the particles. This follows  from Table~\ref{t1}
where the opacities at $\lambda = 1$\,mm are  presented.
This Table contains the results for particles consisting of three
materials (AC1, astrosil and vacuum) or
two materials (AC1 or astrosil and vacuum).
In the first case,
the volume fractions of AC1 and astrosil are equal
($V_{\rm AC1} /V_{\rm total} = V_{\rm astrosil} /V_{\rm total} =
1/2 \, (1 - {\cal P})$) while in the second case the volume fraction
of solid material is $1 - {\cal P}$.
It can be seen that the values of $\kappa$ are generally larger for particles
with a larger fraction of vacuum. This is related
to the decrease of the particle density $\rho_{\rm d}$ which is calculated
as the volume-average quantity.
 As the mass of dust in an object is proportional to $\rho_{\rm d}$
(see Eqs.~(\ref{m}) and (\ref{kap})), the assumption of porous grains
 can lead to considerably smaller mass estimates.
Note that the opacities are larger for more absorbing
carbon particles. A similar effect was  noted by
Quinten et al.~(\cite{qkhm02}) who theoretically studied the wavelength
dependence of extinction of different carbonaceous particles.
They also showed that the far IR extinction was larger
for clusters of spheres and spheroids than for compact spheres.
A very large enhancement of the submm opacities was  found
by Ossenkopf \& Henning~(\cite{oh94}) in the case of pure
carbon aggregates or carbon on silicate grains.

Using Eq.~(\ref{m}) and data from Table~\ref{t1} shows
how the particle porosity and structure
can influence estimates  of dust mass in an object.
The mass ratio estimates can be found in the Rayleigh--Jeans approximation
$$
\frac{M_{\rm d} ({\rm \mbox{compact}})}{M_{\rm d} ({\rm porous})}
=  \frac{\kappa_{\rm porous}(\lambda) \, T_{\rm d, porous}}
    {\kappa_{\rm compact}(\lambda) \, T_{\rm d, compact}}.
$$
With grain temperatures from Fig.~\ref{td-ppp}
and the values of $\kappa$ for composite grains and EMT-Mie theory
(the third column in Table~\ref{t1}) we can find that the ratio
$M_{\rm d} ({\rm \mbox{compact}})/M_{\rm d} ({\cal P} = 0.9)$
lies between $\sim 1.3$ and $\sim 1.5$. If the layered-sphere model
is used (the forth column in Table~\ref{t1}) the ratio increases to
$3.8 - 4.3$. This means that the calculated mass of an object can be reduced
if compact grains are replaced by porous ones.

The ratio of dust masses calculated for two grain models is
$$
\frac{M_{\rm d} ({\rm \mbox{EMT-Mie}})}{M_{\rm d} ({\rm layered \, sphere})}
=  \frac{\kappa_{\rm lay \, sphere}(\lambda) \, T_{\rm d, lay \, sphere}}
    {\kappa_{\rm EMT-Mie}(\lambda) \, T_{\rm d, EMT-Mie}} \approx 2.8.
$$
The numerical value was obtained for particles
consisting of AC1, astrosil and vacuum with ${\cal P} = 0.9$
and the temperature ratio
$T_{\rm d, lay \, sphere}/T_{\rm d, EMT-Mie}=0.85$
as discussed in Sect.~\ref{temp}.
If we consider  particles of the same porosity but consisting of two
materials, the ratio of masses will
be even larger (3.1 for AC1 and 4.3 for astrosil).
Thus,  one can overestimate the mass of an object
by  a factor of 3 or more  if the EMT-Mie model is applied,
since real dust grains in molecular cloud cores
should be very porous and should have non-Rayleigh inclusions.
Another case when the effect can be important is in circumstellar
discs, e.g. Takeushi et al.~(\cite{tcl05}) used $\kappa = 0.3\,{\rm cm^2/g}$
at $\lambda = 1$\,mm for highly porous silicate grains,
which is a good approximation only for particles with small size
inclusions (see Table~\ref{t1}).

\section{Concluding remarks}\label{concl}

We have considered how the porosity of composite cosmic dust grains
can affect their optical properties important for interpretation of
observations of interstellar, circumstellar and cometary dust.
 Two models of particle structure were used.
 Particles of the first kind had well-mixed inclusions
small in comparison with the wavelength,
while those of the second kind consisted of very thin,
cyclically repeating layers.
 Earlier we  showed that the optical properties of such layered particles
are close to those of particles with small and large inclusions
(see Voshchinnikov et al. \cite{vih05}).
 As effective medium theories give reliable results
for particles with small inclusions,
two very different particle structure models can be
simply realized and extensive computations can be performed.

For both models, we studied
how an increase of the volume fraction of vacuum could change
the extinction efficiencies at different wavelengths,
temperature of dust grains, profiles of the IR silicate bands and
dust millimeter opacities.
 It is found that the models begin to differ
essentially when the porosity exceeds $\sim 0.5$.
 This difference appears as lower temperatures (Sect.~\ref{temp}),
shifted central peaks of the silicate bands
(Sect.~\ref{ir_b}) and larger millimeter opacities
(Sect.~\ref{opa})
for layered particle model in comparison with that based
on EMT calculations.
 The latter model also requires larger dust-phase abundances
than the layered model (Sect.~\ref{st_ext})
to produce the same interstellar extinction.

The assumption that interstellar particles have only small size inclusions
looks to some extent artificial
(excluding, of course, the case of special laboratory samples).
 Therefore, we believe that the layered sphere model well describing
light scattering by very porous quasispherical particles with
inclusions of different sizes
should find wide applications in interpretation of different phenomena.
 In particular, this model has good perspectives
to explanation of the flat interstellar extinction observed
in the near-IR part of spectrum (Sect.~\ref{ir_ext}) and
variations of the shape of the silicate feature detected in spectra
of T Tau and Herbig Ae/Be stars
(the results will be in a subsequent next paper).

\acknowledgements{
We are grateful to Walter Wegner for the possibility to use unpublished data
and to Bruce Draine for  comments on an earlier version of the paper.
We are also grateful to the referee Michael Wolff and scientific editor
Anthony Jones for useful suggestions.
NVV acknowledges the hospitality of the Max-Planck-Institut f\"ur Astronomie
where this work was finished.
The work was partly supported by grant 1088.2003.2 of the President of the
Russian Federation for leading scientific schools.
}


\end{document}